# DEPENDENCE OF THE 27-DAY VARIATION OF COSMIC RAYS ON THE GLOBAL MAGNETIC FIELD OF THE SUN

R. Modzelewska[a] and M.V. Alania[a, b]

*(a) Institute of Math. and Physics of University of Natural Sciences and Humanities, 08-110 Siedlce, Poland*
*(b) Institute of Geophysics of Tbilisi state University, Tbilisi, Georgia*

## ABSTRACT

We show that the higher range of the heliolongitudinal asymmetry of the solar wind speed in the positive polarity period (A>0) than in the negative polarity period (A<0) is one of the important reasons of the larger amplitudes of the 27-day variation of the galactic cosmic ray (GCR) intensity in the period of 1995-1997 (A>0) than in 1985-1987 (A<0). Subsequently, different ranges of the heliolongitudinal asymmetry of the solar wind speed jointly with equally important corresponding drift effect are general causes of the polarity dependence of the amplitudes of the 27-day variation of the GCR intensity. At the same time, we show that the polarity dependence is feeble for the last unusual minimum epoch of solar activity 2007-2009 (A<0); the amplitude of the 27-day variation of the GCR intensity shows only a tendency of the polarity dependence. We present a three dimensional (3-D) model of the 27-day variation of GCR based on the Parker's transport equation. In the 3-D model is implemented a longitudinal variation of the solar wind speed reproducing in situ measurements and corresponding divergence-free interplanetary magnetic field (IMF) derived from the Maxwell's equations. We show that results of the proposed 3-D modeling of the 27-day variation of GCR intensity for different polarities of the solar magnetic cycle are in good agreement with the neutron monitors experimental data. To reach a compatibility of the theoretical modeling with observations for the last minimum epoch of solar activity 2007-2009 (A<0) a parallel diffusion coefficient was increased by ~ 40%.

## INTRODUCTION

Richardson et al. (1999) found that the size of the recurrent cosmic ray modulations is larger when A>0 than during A<0. Later, it was demonstrated that the amplitudes of the 27-day variation of the GCR intensity obtained from neutron monitors data and theoretical modeling are greater in the minimum epochs of solar activity for the A>0 than for the A<0 (Alania et al., 2001a; 2001b; Gil and Alania 2001; Vernova et al., 2003; Iskra et al., 2004). Solving numerically a three-dimensional model of GCR transport, Kota and Jokipii (2001) showed that larger recurrent variation of the proton flux can be expected when A>0 than when A<0. Burger and Hitge (2004) and Burger et al. (2008) developed a hybrid model of the IMF, and suggested that the Fisk heliospheric magnetic field can explain several properties of the 27-day cosmic ray variation. Unfortunately, there are problems concerning the existence of the Fisk's type heliospheric magnetic field (Fisk, 1996) in the minimum epoch of solar activity (Roberts et al., 2007). Modzelewska et al. (2006), Alania et al. (2008a) demonstrated that the heliolongitudinal distribution of the phase of the 27-day variation of solar wind speed has a clear maximum in 1995-1997 (A>0), but it remains obscure for 1985-1987 (A<0). More, the phase distribution of the 27–day variation of the solar wind speed shows that the heliolongitudes with the more stable long-lived recurrent conditions are observed in the A>0 than in the A<0 solar magnetic polarity periods for the earlier minimum epochs of solar activity. This phenomenon can be considered as a source of the larger 27-day variation of the solar wind speed in 1995-1997 (A>0) than in 1985-1987 (A<0).

We believe that for two particular cases the higher range of the heliolongitudinal asymmetry of the solar wind speed in the A>0 than in the A<0 is a general cause of the larger amplitudes of the 27-day variation of the GCR intensity in 1995-1997 (A>0) than in 1985-1987 (A<0). Recently, for the first time, it was demonstrated that also the amplitudes of the 27-day variation of the GCR anisotropy at solar minimum are greater when A>0 than when A<0 (Alania et al., 2005; 2008a; 2008b; Gil et al., 2005).

However, it must be underlined that in papers (Gil and Alania, 2001; Alania et al., 2005; Gil et al., 2005; Alania et al., 2008a; 2008b; Kota and Jokipii, 2001; Burger and Hitge, 2004; Burger et al., 2008), in order to explain results of Richardson et al. (1999), the general attention was paid to the drift effect and the role of recurrent changes of the solar wind speed in different magnetic polarities of the Sun, which should be a vital (Alania et al., 2008a, Gil et al., 2008), was not considered. Comparison of observations from Ulysses first three orbits demonstrated that the 3-D structure of the solar wind varies dramatically over the solar cycle (McComas et al., 2006). They suggest that this difference might be a regular feature of the opposite solar magnetic orientation phases of the full ~22-year Hale cycle. Lario and Roelof (2007) reported that the long-lived and well defined ~26-day recurrent intensity enhancements is observed in the minimum epoch of solar activity when A>0 during the first (1992 - 1994) southern excursion of the Ulysses. In contrast to this, during the third (2005 - 2007) southern excursion of the Ulysses, taking place also in low solar conditions but in A<0, observations showed more variable structure in the solar wind stream. Dunzlaff et al. (2008) suggest that this might be caused by the difference in the coronal hole structures between the cycles 22 (1986-1996) and 23 (1996-2008) due to a large, stable coronal hole structure, which is present during cycle 22, but not in cycle 23. It was demonstrated that there is a consistent and generally significant relationship between cosmic ray recurrent variation and solar wind speed, especially for A>0 (Modzelewska et al., 2006; Singh and Badruddin, 2007; Alania et al., 2008a; Gupta and Badruddin, 2009). Before, we found (Modzelewska et al., 2006; Alania et al., 2008a) that the 27-day variation of the solar wind speed is more regular in the positive polarity period 1995-1997 (A>0), than in 1985-1987 (A<0). At the same time, other parameters (interplanetary magnetic field, sunspot number, heliospheric current sheet tilt etc.) do not show any regular changes as the solar wind speed. Bearing in mind a significance of the drift for different polarity epochs, we ascribe a distinction of the amplitudes of the 27 –day variation of GCR intensity in particular periods of 1995-1997 and 1985-1987 to the valuable role of the difference in magnitude of the 27-day variation of the solar wind speed.

The main purpose of this paper is: (1) to study a relationship of the 27-day variations of the GCR intensity and solar wind speed based on experimental data over the course of three consecutive minimum epochs of solar activity with the opposite solar magnetic polarities, (2) to develop a three dimensional (3-D) model of the 27-day variation of the GCR intensity implementing: (I) a longitudinal variation of the solar wind speed reproducing in situ measurements, and (II) a corresponding divergence-free interplanetary magnetic field derived from the Maxwell's equations.

DATA ANALYSIS

To study relationships of the 27-day variations of the solar wind speed and the GCR intensity over the course of three successive minimum epochs of solar activity with opposite polarities (1985-1987, A<0; 1995-1997, A>0 and 2007-2009, A<0) we use data of the solar wind speed from SPIDR (http://spidr.ngdc.noaa.gov/spidr/) and OMNI (http://omniweb.gsfc.nasa.gov/index.html) data sets, and Kiel neutron monitor data

(http://cr0.izmiran.rssi.ru/kiel/main.htm, http://www.nmdb.eu). Unfortunately, the solar wind data in the mid 1980's, provided by IMP 8, which spent about 40% of each ~13 day geocentric orbit inside the Earth's bow shock, are incomplete. We use linear Lagrange interpolation to fulfill existing data gaps for 1985-1987.

In statistical investigation relating the 27-day variations of the solar wind speed and the GCR intensity over the course of three successive minimum epochs of solar activity with opposite polarities the main assumptions is that the longitudinal structures are stable over several years, and the Bartels period (27 days) is adequate for that. Therefore for analyses of the experimental data we employ a method of superposed epoch. Using harmonic analysis method we determine the best fit sine wave to each Bartels rotation period. Each epoch corresponds to the superimposed 27-day group of the 27-day variation of the solar wind speed and recurrent GCR intensity variation. In Fig.1a, b, c are presented temporal changes of superimposed solar wind speed (points) fitting by a 27-day harmonic wave (dashed lines) in 1985-1987 (A<0) (Fig. 1a), 1995-97 (A>0) (Fig. 1b) and 2007-2009 (A<0) (Fig. 1c). Fig. 1 shows that the range of changes of the 27-day variation of the solar wind speed is greater in the period of 1995-1997 (A>0), than in 1985-1987 (A<0) (Gil et al., 2008). Similar results are obtained in recently published paper (Gupta and Badruddin, 2009). Nevertheless, there remains some uncertainty due to uncompleted data of solar wind speed for the period of 1985-1987 (A<0). However, in some degrees we could avoid this uncertainty by analyzing data of the geomagnetic AA index. The point is that AA index unquestionably well correlate with the solar wind speed (e.g. Rangarajan and Barreto, 2000; Laptukhov and. Laptukhov, 2010). In Fig. 2a, b, c are presented temporal changes of superimposed AA index (points) fitting by a 27-day harmonic wave (dashed lines) for three successive minimum epochs of solar activity with different polarities, 1985-1987 (A<0), 1995-1997 (A>0) and 2007-2009 (A<0). Fig.2a shows that, the 27-day variation of the AA index is hardly recognizable in the period 1985-1987 (A<0), while in the period 1995-1997 (A>0) and 2007-2009 (A<0) it is clearly seen. So, the changes of the AA index are similar as the changes of the solar wind velocity in the periods of 1985-1987 (A<0), 1995-1997 (A>0) and 2007-2009 (A<0). Therefore, it is possible to accept that the range of regular changes of solar wind speed, e.g. the first harmonic of the 27-day variation, is larger in the period of 1995-1997 (A>0), than in the period of 1985-1987 (A<0). It must be emphasized that in 2007-2009 (A<0) the magnitude of the 27-day variation of the solar wind speed remains almost at the same level as in 1995-1997 (A>0), but the range of observed solar wind speed is larger with great second (14 days) and third (9 days) harmonics.

For further analyzes we calculated the amplitudes and phases of the 27-day variations of the GCR intensity and solar wind speed based on daily data using the harmonic analyses method during each Bartels rotation period. To study a heliolongitudinal asymmetry of the solar wind velocity and of the GCR intensity we use phase histograms method. In Fig. 3 and 4 are presented the histograms of the phases of the 27-day variation of the solar wind speed (Fig. 3) and GCR intensity obtained by Kiel neutron monitor (Fig. 4) for A<0 (1985-1987; 2007-2009) and 1995-1997 (A>0) consecutive minimum epochs of solar activity. In the histograms (Figs. 3 and 4), nine intervals of phases of the 27-day variations with the step of $40^0$ of the heliolongitudes were created: $0-40^0$, $41-80^0$, $81-120^0$, …, and $321-360^0$. Each column (frequency n of given range of phases) is considered as a Poisson's statistics and the error bar is calculated as $\sqrt{n}$. The histogram of the phases of the 27-day variation of the solar wind speed in 1995-1997 (Fig. 3b) shows a tendency of higher concentration of the phases about $\sim 320^0 - 360^0$. At same time, in 1985-1987 (A<0) (Fig. 3a), the heliolongitudes with the long-lived recurrent conditions have generally the same location as in 1995-1997 (A>0), but with higher dispersion. In recent minimum epochs of solar activity, 2007-2009 (A<0) (Fig. 3c) the histogram of the phase distribution of 27-day variation of solar wind speed shows almost the same location of concentration as in 1995-1997 (A>0). The characters of the presented histograms for opposite polarities result to the ~ two times greater average magnitude of the heliolongitudinal asymmetry of the solar wind speed in the period of 1995-1997 (A>0) and 2007-2009 (A<0) than in 1985-1987 (A<0). In Fig.4 are presented the histograms of the phases of the 27-day variation of the GCR intensity obtained by Kiel neutron monitor data (Fig 4a) for the 1985-1987 (A<0), (Fig 4b) for the 1995-1997 (A>0) and (Fig 4c) for the 2007-2009 (A<0). The histogram of the phases of the 27-day variation of the GCR intensity in 1995-1997 (A>0) (Fig. 4b) shows a tendency of higher concentration (maximum) of the phase distribution about ~$120^0$-$240^0$ of heliolongitudes, and the histogram (Fig. 4a) for 1985-1987 (A<0) shows some tendency of concentration (maximum) of the phase distribution in the same region as for A>0, but with larger dispersion. In recent minimum epoch of solar activity 2007-2009 (A<0) the histogram of the phases of the 27-day variation of the GCR intensity (Fig. 4c) shows a peak -maximum at ~200-$240^0$ of heliolongitudes. Correlation coefficient between the phase distributions of cosmic rays and solar wind in 1995-1997 (A>0) is negative and significant (-0,68 ± 0,16), while for 1985-1987 and 2007-2009 (A<0) it is negative, as well, but insignificant (-0,22 ± 0,22 and -0,44 ± 0,21, respectively). At first approximation for all three minimum epochs of 1985-1987 (A<0), 1995-1997 (A>0) and 2007-2009 (A<0) the heliolongitudinal location of the sources of the 27-day variations are the same. It confirms an existence of the long-lived heliolongitudal asymmetry of solar activity and solar wind (Vitinsky, 1983; Neugebauer et al., 2000; Berdyugina and Usoskin, 2003).

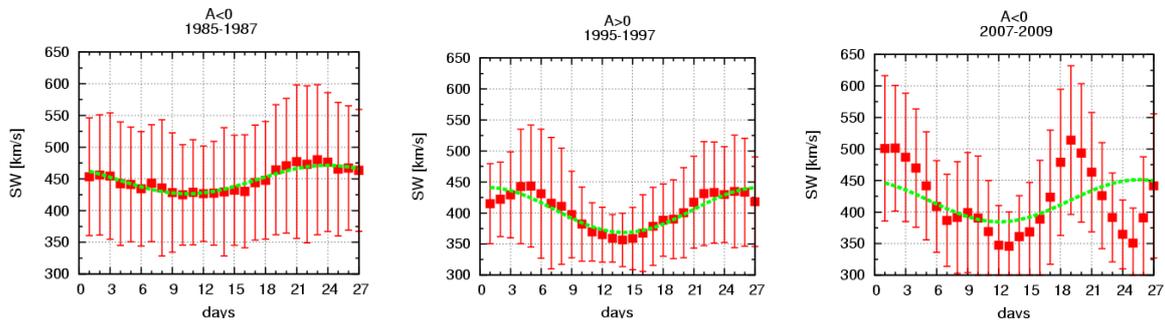

Fig. 1abc. Temporal changes of superimposed solar wind speed (points) fitting by a 27-day harmonic wave (dashed line) for 1985-1987 (A<0), 1995-1997 (A>0) and 2007-2009 (A<0) three successive minimum epochs of solar activity.

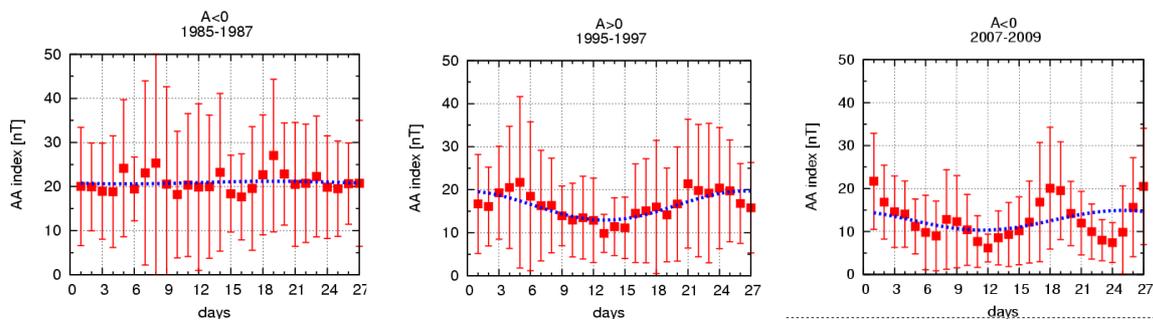

Fig. 2abc. Temporal changes of superimposed AA index (points) fitting by a 27-day harmonic wave (dashed line) for 1985-1987 (A<0), 1995-1997 (A>0) and 2007-2009 (A<0) three successive minimum epochs of solar activity.

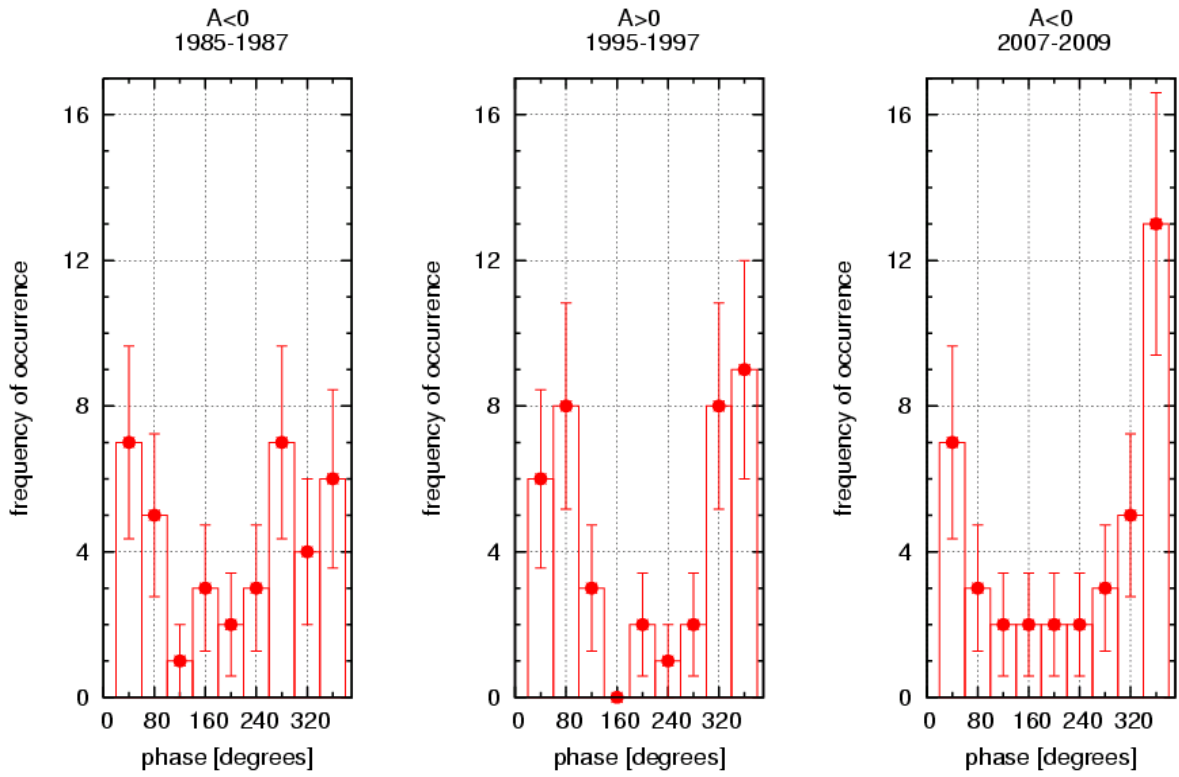

Fig. 3abc. Histograms of the phases of the 27-day variation of solar wind speed for (a) 1985-1987 (A<0), (b) 1995-1997 (A>0) and (c) 2007-2009 (A<0) three successive minimum epochs of solar activity (Alania et al., 2008a; Alania et al., 2008b).

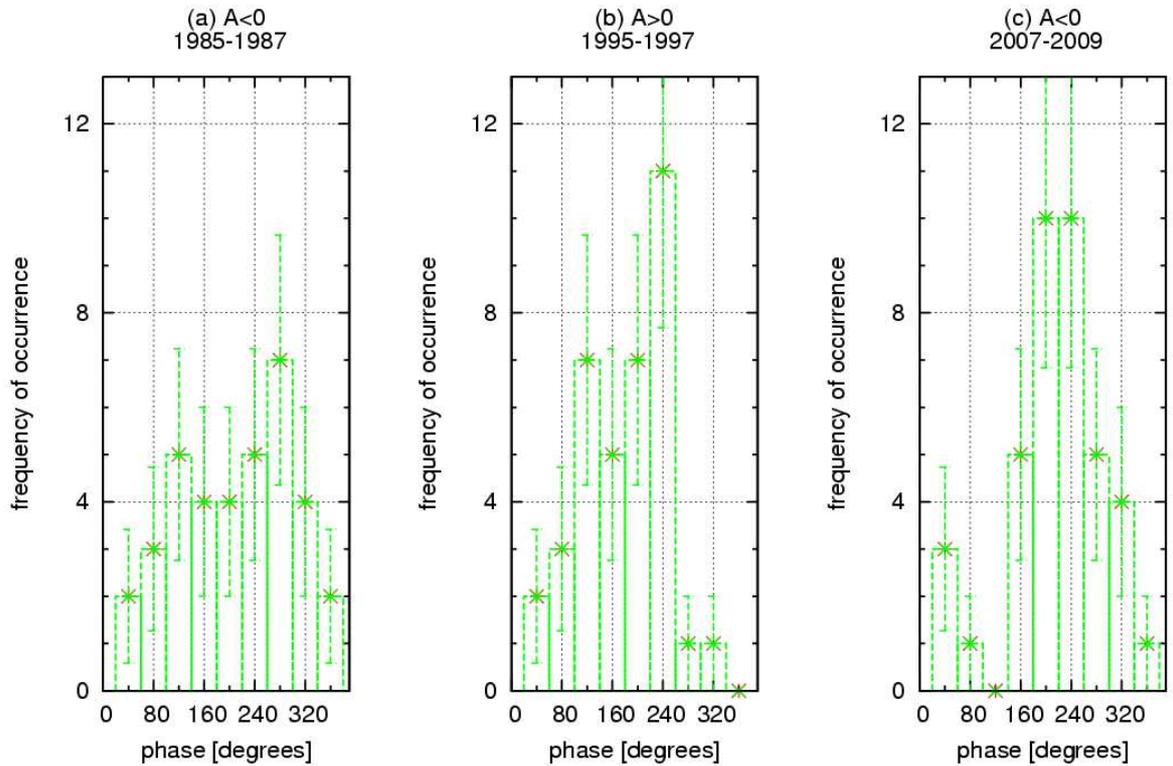

Fig. 4abc Histograms of the phases of the 27-day variation of the GCR intensity obtained by Kiel neutron monitor for (a) 1985-1987 (A<0), (b) 1995-1997 (A>0) and (c) 2007-2009 (A<0) three successive minimum epochs of solar activity.

The recurrent processes related with the certain heliolongitudes setting usually by the coronal holes in the minimum epochs and by the heliolongitudinally asymmetric solar activity generally, can be considered as a source of the long-lived 27-day variation of the solar wind speed. The greater amplitudes of the 27–day variation of the GCR intensity in 1995-1997 (A>0) than in 1985-1987 (A<0) observed by neutron monitors (Richardson et al., 1999; Gil and Alania, 2001), and the similar results obtained for the anisotropy (Alania et al., 2005; Gil et al., 2005; Alania et al., 2008a) are related with the valuable difference in magnitude of the 27-day variation of solar wind speed and by an equally important effect of drift. In recent relatively long-lasting minimum epoch 2007-2009 (A<0) of solar activity (the ending part of the solar cycle #23) the background solar activity parameters were appreciably different in comparison with the previous solar minima. The sun was extremely quiet, with almost no sunspot on its surface (e.g. Smith, 2011). At the same time mean value of the strength B of the regular IMF in the period of 2007-2009 was record-low (~2.5 nT) in comparison with the previous minimum epochs, (~4 nT in 1985-1987 and ~3.4 nT in 1995-1997). All at once large-scale configuration of 3-D heliosphere were well organized with the stable over several years heliolongitudinal structures. It is especially pronounced in the behavior of the solar wind speed. The magnitude of the 27-day variation of the solar wind speed remains almost at the same level as in 1995-1997 (A>0), but the range of observed solar wind speed is larger with the significant second (14 days) and third (9 days) harmonics. It is well known that the diffusion coefficient K of the GCR particles, depends, among other parameters, on the magnitude B of the regular IMF, as K∝1/B, implying, in recent minimum epoch of solar activity, relatively high parallel diffusion coefficient, which gives a natural reason for the higher galactic cosmic ray intensities that were observed by neutron monitors (e. g. Moraal and Stoker, 2010) and space probes (e. g. Mewaldt et al., 2010), as well. We suppose that an unique character of the spatial large-scale structure of the solar wind speed with its stable heliolongitudinal asymmetry preferentially pronounces in behavior of the 27-day variation of the GCR intensity in the long-lasting recent minimum epoch of extremely low solar activity.

THEORETICAL MODELING

Solar magnetic cycle dependence of the recurrent variations of cosmic rays first reported by Richardson et al. (1999) aimed to be explained in papers (Gil and Alania, 2001; Alania et al., 2005; Gil et al., 2005; Alania et al., 2008a, 2008b; Kota and Jokipii, 2001; Burger and Hitge, 2004; Burger et al., 2008) by predictions of drift theory. At the same time the role of recurrent variation of the solar wind speed in different magnetic polarities was not considered. The 3-D models with heliolongitudinally dependent solar wind velocity with consequent changes of the frozen-in IMF were presented by Kota and Jokipii (1991; 2001) followed after the first 3-D model (Kota and Jokipii, 1983). Model of Kota and Jokipii (1991) was the first global simulations of the modulation of GCR by 3-D solar wind with corotating interaction regions (CIR), though there was not considered an expected difference of the amplitudes of the 27-day variation of the GCR intensity in the A>0 and A<0 polarity periods. Later, Kota and

Jokipii (2001) developed a new time dependent 3-D model including multiple heliospheric neutral sheet (HNS) characterizing high solar activity to cover more complex structures of the HNS and CIRs. It was demonstrated that the recurrent variation (27-days) of the GCR intensity is larger in the A>0 polarity epoch than in the A<0 epoch in qualitative agreement with the results of Richardson et al. (1999). 3-D models developed by Kota and Jokipii (1983; 1991; 2001) show that in the global modulation model of GCR, the large-scale structure is controlled by drift effects in conjunction with diffusion, convection and energy change, but the small-scale structure is caused by diffusive effects in the transient and corotating structures. The motivation for the current model is our findings (previous section) that the magnitude of the heliolongitudinal asymmetry of the solar wind speed is larger in the periods of 1995-1997 (A>0) and 2007-2009 (A<0) than in 1985-1987 (A<0) in the minimum and near minimum epochs of solar activity. So, to construct an appropriate model of the 27-day variation of the GCR intensity based on Parker's (1965) transport equation, the spatial dependences of the solar wind speed $V$ and the IMF $B$ must be taken into account. However, it is quite difficult problem, because the divergence free condition ($\nabla \cdot B = 0$) of the strength B of the IMF should be kept for the spatially dependent solar wind speed.

To investigate theoretically a solar magnetic cycle dependence of the 27-day variation of cosmic rays we use a steady state, three-dimensional Parker's transport equation (Parker 1965):

$$\nabla_i \cdot (K_{ij}^S \cdot \nabla_j f) - (v_{d,i} + V_i) \cdot \nabla_j f + \frac{1}{3}(\nabla_i \cdot V_i)\frac{\partial f}{\partial \ln R} = 0 \tag{1}$$

Where $f$ and $R$ are omnidirectional distribution function and rigidity of cosmic ray particles, respectively; $V_i$ is the solar wind speed, $t$ is time, $v_{d,i}$ is the drift velocity and $K_{ij}^S$ is the symmetric part of anisotropic diffusion tensor $K_{ij}$.

The drift velocity is implemented as $<v_{d,i}> = \partial K_{ij}^A / \partial x_j$ (Levy, 1976; Jokipii et al., 1977) where $K_{ij}^A$ is the antysymmetric part of anisotropic diffusion tensor ($K_{ij} = K_{ij}^S + K_{ij}^A$) of GCR.

Our aim in this paper is to compose a model of the 27-day variation of the GCR intensity for the solar wind speed depending on heliolongitude reproducing in situ measurements for different polarities of the solar magnetic cycle.

In this model we assume that the stationary 27-day variation of the GCR intensity is caused by the heliolongitudinal asymmetry of the solar wind speed and influenced by drift in different polarity epochs. For this purpose we included in Eq. (1) the changes of the radial solar wind speed (dashed curves in Fig. 1a, b, c) corresponding to in situ measurements for three consecutive minimum epochs of solar activity with opposite polarities. The measured in situ radial speeds were approximated by the formulas:

in 1985-1987 (A<0), $V_r = V_0(1 + \alpha \sin(0.86 - \varphi))$, $\alpha = 0.05$ (2a)

in 1995-1997 (A>0), $V_r = V_0(1 + \alpha \sin(1.42 + \varphi))$, $\alpha = 0.09$ (2b)

in 2007-2009 (A<0), $V_r = V_0(1 + \alpha \sin(1.9 + \varphi))$, $\alpha = 0.08$ (2c)

where $\varphi$ is the heliographic longitude.

To solve Eq. (1) with the longitudinal variation of solar wind speed, it is necessary to implement the corresponding divergence-free IMF. For this purpose the Maxwell's equations should be solved for the solar wind speed reproducing in situ measurements (expressions (2a), (2b) and (2c)) for particular polarity periods.

Maxwell's equations describing the divergence free condition for the strength B of the IMF frozen in the electrically high conductive solar wind moving with the speed V (e. g. Parker, 1963; Jackson, 1998) are:

$$\begin{cases} \dfrac{\partial B}{\partial t} = \nabla \times (V \times B) & (3a) \\ \nabla \cdot B = 0 & (3b) \end{cases}$$

The system of scalar equations for the components $(B_r, B_\theta, B_\varphi)$ of the IMF and components $(V_r, V_\theta, V_\varphi)$ of the solar wind speed can be rewritten in corotating frame (attached to the rotating Sun) in the heliocentric spherical $(r, \theta, \varphi)$ coordinate system, as:

$$\begin{cases} \dfrac{\partial B_r}{\partial t} = \dfrac{1}{r^2 \sin\theta} \left[ \dfrac{\partial}{\partial \theta}[(V_r B_\theta - V_\theta B_r)r\sin\theta] - \dfrac{\partial}{\partial \varphi}[(V_\varphi B_r - V_r B_\varphi)r] \right] & (4a) \\ \dfrac{\partial B_\theta}{\partial t} = \dfrac{1}{r \sin\theta} \left[ \dfrac{\partial}{\partial \varphi}(V_\theta B_\varphi - V_\varphi B_\theta) - \dfrac{\partial}{\partial r}[(V_r B_\theta - V_\theta B_r)r\sin\theta] \right] & (4b) \\ \dfrac{\partial B_\varphi}{\partial t} = \dfrac{1}{r} \left[ \dfrac{\partial}{\partial r}[(V_\varphi B_r - V_r B_\varphi)r] - \dfrac{\partial}{\partial \theta}(V_\theta B_\varphi - V_\varphi B_\theta) \right] & (4c) \\ \dfrac{1}{r}\dfrac{\partial}{\partial r}(r^2 B_r) + \dfrac{1}{\sin\theta}\dfrac{\partial}{\partial \theta}(\sin\theta B_\theta) + \dfrac{1}{\sin\theta}\dfrac{\partial}{\partial \varphi}B_\varphi = 0 & (4d) \end{cases}$$

We consider a model of the IMF when $V_\theta$ and $B_\theta$ equal zero. This assumption straightforwardly leads (from Eq. (4a)) to the relationship between $B_r$ and $B_\varphi$, as, $B_\varphi = B_r \dfrac{V_\varphi}{V_r}$, where $V_\varphi = -\Omega r \sin\theta$ is the corotational speed. Therefore, in this case our goal is to find radial and azimuthal components of the stationary IMF for the variable solar wind speed reproducing in situ measurements. Taking into account the expressions for $B_\varphi$ and $V_\varphi$ we obtain from Eq. (4d) an equation with respect to the radial component $B_r$:

$$A_1 \dfrac{\partial B_r}{\partial r} + A_2 \dfrac{\partial B_r}{\partial \varphi} + A_3 B_r = 0 \qquad (5)$$

The coefficients $A_1$, $A_2$ and $A_3$ depend on the radial $V_r$ and heliolongitudinal $V_\varphi$ components of the solar wind speed V. Equation (5) was reduced to the algebraic system of equations using a difference scheme method and then solved by the method of iteration. Details of the numerical solution of Eq. (5) are discussed in recently published paper (Alania et al., 2010).

In Parker's transport equation we included corresponding $B_r$ and $B_\varphi$ components and the magnitude $B = \sqrt{B_r^2 + B_\varphi^2}$ of the IMF obtained from the numerical solution of Eq. (5) with the heliolongitudinally dependent solar wind speed given by expressions (2a), (2b) and (2c) for corresponding solar magnetic polarities. The heliospheric magnetic field vector $\overline{B}$ is written (Jokipii and Kopriva, 1979):

$$\overline{B} = \left(1 - 2H\left(\theta - \frac{\pi}{2}\right)\right)\left(B_r \overline{e_r} + B_\varphi \overline{e_\varphi}\right)$$

where H is the Heaviside step function changing the sign of the global magnetic field in each hemisphere. Radial component $B_r$ is obtained from the numerical solution of Eq. (5), azimuthal component $B_\varphi$ is calculated according to the expression $B_\varphi = B_r \cdot \frac{V_\varphi}{V_r}$, where $V_\varphi$ is the corotational speed and $V_r$ radial solar wind speed given by expressions (2a), (2b) and (2c) correspondingly, for different polarities.

The flat neutral sheet drift was taken into account according to the boundary condition method (Jokipii and Kopriva, 1979), when the delta function at the solar equator is a consequence of the abrupt change in sign of the IMF. Implementation of the interplanetary magnetic field obtained from the numerical solution of Eq. (5) in Parker's transport equation (1) is done through the spiral angle $\psi = \arctan\left(-\frac{B_\varphi}{B_r}\right)$ in anisotropic diffusion tensor of GCR particles and through the magnitude $B = \sqrt{B_r^2 + B_\varphi^2}$ of the IMF in the ratios $\alpha = \frac{\kappa_\perp}{\kappa_\parallel}$ and $\alpha_1 = \frac{\kappa_d}{\kappa_\parallel}$ of the perpendicular $\kappa_\perp$ and drift $\kappa_d$ diffusion coefficients to the parallel $\kappa_\parallel$ diffusion coefficient. The parallel diffusion coefficient $\kappa_\parallel$ changes versus the spatial spherical coordinates $(r, \theta, \varphi)$ and rigidity R of GCR particles as, $\kappa_\parallel = \kappa_0 \kappa(r) \kappa(R)$, where $\kappa_0 = \frac{\lambda_0 v}{3} = 2 \cdot 10^{22} \, cm^2/s$, $v$ is the speed of GCR particles; $\kappa(r) = 1 + \alpha_0 r$, $\alpha_0 = 0.5$; $\kappa(R) = (R/1GV)^{0.5}$. So, the parallel diffusion coefficient for the GCR particles of rigidity 10 GV equals, $\kappa_\parallel \approx 10^{23} \frac{cm^2}{s}$ at the Earth orbit. The ratios $\alpha = \frac{\kappa_\perp}{\kappa_\parallel}$ and $\alpha_1 = \frac{\kappa_d}{\kappa_\parallel}$ have a standard form: $\alpha = \frac{\kappa_\perp}{\kappa_\parallel} = (1 + \omega^2 \tau^2)^{-1}$ and $\alpha_1 = \frac{\kappa_d}{\kappa_\parallel} = \omega\tau(1 + \omega^2 \tau^2)^{-1}$, where $\omega\tau = 300 \, B \lambda_0 R^{-1}$, B is the strength of the IMF changing vs. radial distance and $\lambda_0$ is the transport free path of GCR particles.

The kinematical model of the IMF with variable solar wind speed has some limitations, especially it would be applied until some radius, while at large radii the faster wind would overtake the previously emitted slower one. To avoid an intersection of the IMF lines the heliolongitudinal asymmetry of the solar wind speed takes place only up to the distance of ~ 8 AU and then $V = 400 \, km/s$ throughout the heliosphere. Behind 8 AU the standard Parker's field is applied. Equation (1) was solved numerically using a difference scheme method and then solved by the Gauss-Seidel method of iteration as in our papers published elsewhere

(e.g., Alania, 2002; Iskra et al., 2004; Modzelewska et al., 2006; Wawrzynczak and Alania, 2008; Alania et al., 2010).

Results of the solution of Eq. (1) are presented in Fig. 5. In Fig. 5 are presented the expected amplitudes of the 27-day variation of the GCR intensity calculated based on the solutions of Eq. (1) with longitudinally dependent solar wind speed for 1985-1987 (A<0), 1995-1997 (A>0) and 2007-2009 (A<0) consecutive minimum epochs of solar activity (red filled cycles). In this model (further in this paper called model I) the parallel diffusion coefficient for the GCR particles of rigidity 10 GV equals $\kappa_{\parallel} \approx 10^{23} \frac{cm^2}{s}$ at the Earth orbit. Also, in this Fig. 5 are presented the amplitudes of the 27-day variation of the GCR intensity calculated by Kiel neutron monitor data (crosses with error bars) for two A>0 periods (1975-1977 and 1995-1997), and for three A<0 periods (1965-1967, 1985-1987 and 2007-2009) (Alania et al., 2008a). Fig. 5 shows that for 1985-1987 (A<0) and 1995-1997 (A>0) the amplitudes expected from the 3-D model I of the 27-day variation of the GCR intensity and from the Kiel neutron monitor experimental data coincide in the scope of uncertainties keeping a clear polarity dependence, while for the 2007-2009 (A<0) period the expected amplitude of the 27-day variation is larger than the amplitude found by observations. Nevertheless, observations of solar wind parameters near the Earth orbit in the last minimum epoch of solar activity imply that the estimated parallel diffusion coefficient for cosmic ray transport was greater in 2009 than in the 1997-1998 solar minimum epoch for ~ 20% according to (Moraal and Stoker, 2010), and ~40%- according to (Mewaldt et al., 2010). So, to reach a compatibility of the theoretical modeling of the 27-day variation of the GCR intensity with the experimental data for the last minimum epoch of solar activity (2007-2009, A<0), we increase the parallel diffusion coefficient $\kappa_{\parallel} \approx 10^{23} \frac{cm^2}{s}$ by 20% ($\kappa_{\parallel} \approx 1.2 \times 10^{23} \frac{cm^2}{s}$, model II) , and by 40% ($\kappa_{\parallel} \approx 1.4 \times 10^{23} \frac{cm^2}{s}$ -model III). Results of the solutions of the model II (square) and model III (triangle)- the expected amplitude of the 27-day variation of the GCR intensity for 2007-2009 (A<0 ) are presented in Fig. 5. Fig. 5 shows that when diffusion coefficient is about 20% greater (model II), the expected amplitude of the 27-day variation of GCR intensity (square) is on the level of 1995-1997 (A>0), but when diffusion coefficient is about 40% greater (model III), there is a tendency (triangle) of a polarity dependence of the 27–day variation of the GCR intensity, as it is shown by experimental data. Additionally, it should be note that an uniqueness of the last minimum epoch of solar activity (2007-2009, A<0) contributes peculiarly in the behavior of the amplitude of the 27-day variation of GCR intensity, and its polarity dependence of the Sun's global magnetic field is shading.

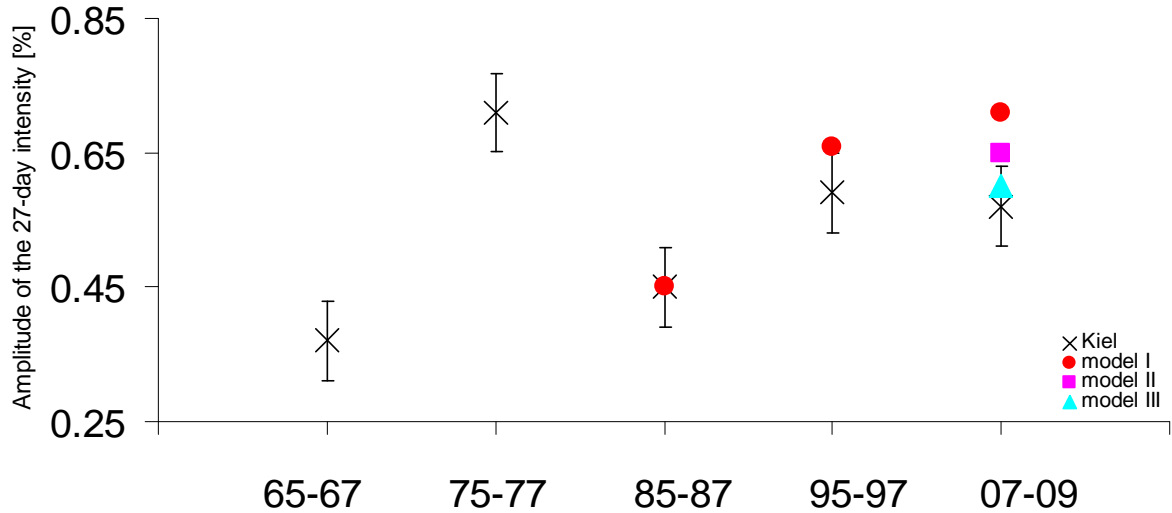

Fig. 5. Amplitudes of the 27-day variation of the GCR intensity obtained by Kiel neutron monitor (crosses) for A>0 (1975-77 & 1995-1997) and A<0 (1965-67, 1985-1987 & 2007-2009) polarity periods (Alania et al., 2008a). Red filled cycles (model I), square (model II) and triangle (model III) designate expected amplitudes of the 27-day variation for GCR particles of rigidity 10 GV obtained based on the solution of Eq. (1) for three consecutive minimum epochs of solar activity.

Thus, one can conclude that the proposed 3-D model of the 27-day variation of the GCR intensity gives a possibility to be used for different polarity periods of solar magnetic cycle with implementing in situ measurements of solar wind speed and corresponding IMF components derived from Maxwell's divergence free equation.

SUMMARY AND CONCLUSIONS

1. The recurrent processes related with the certain heliolongitudes setting generally by the coronal holes in the minimum epochs and partly by the heliolongitudinally asymmetric solar activity is considered as a source of the long-lived 27-day variation of the solar wind speed. We show that the heliolongitudes with the stable long-lived recurrent conditions are observed in sequence minimum epochs of 1985-1987 (A<0), 1995-1997 (A>0) and 2007-2009 (A<0) with some peculiarities. Particularly, the higher range of the heliolongitudinal asymmetry of the solar wind speed in the A>0 than in the A<0 is one of the general cause of the larger amplitudes of the 27-day variation of the GCR intensity for the period of 1995-1997 (A>0) than in 1985-1987 (A<0).
2. In recent relatively long-lasting minimum epoch 2007-2009 (A<0) of solar activity (the ending part of the solar cycle #23) the background solar activity parameters were appreciably different in comparison with the previous solar minima. The sun was extremely quiet, at the same time mean value of the strength B of the regular IMF was record-low (~2.5 nT) in comparison with the previous minimum epochs. All at once large-scale configuration of 3-D heliosphere were well organized with the stable over several years heliolongitudinal structures. It is especially pronounced in the behavior of solar wind speed. The magnitude of the 27-day variation of the solar wind speed remains almost at the same level as in 1995-1997 (A>0), but the range of observed solar wind speed is larger with the significant second (14 days) and third (9 days) harmonics. It is well known that the diffusion coefficient K of the GCR particles, depends, among other parameters, on the magnitude B of the regular IMF, as K∝1/B, implying, in recent minimum epoch of solar activity, relatively high parallel diffusion coefficient, which gives a natural reason for the higher galactic cosmic ray intensities that were observed by neutron monitors (e. g. Moraal and Stoker, 2010) and space probes (e. g. Mewaldt et al., 2010), as well. We suppose that an unique character of the spatial large-scale structure of the solar wind speed with its stable heliolongitudinal asymmetry preferentially pronounces in behavior of the 27-day variation of the GCR intensity in the last minimum epoch of solar activity (2007-2009;A<0); the amplitude of the 27-day variation of the GCR intensity only shows a tendency of the polarity dependence. It seems that an uniqueness of the last minimum epoch of solar activity (2007-2009, A<0) contributes peculiarly in the behavior of the amplitude of the 27-day variation of GCR intensity, and its polarity dependence of the Sun's global magnetic field is shading.
3. We present a 3-D model of the 27-day variation of the GCR intensity based on the Parker's transport equation with the longitudinal variation of the solar wind speed reproducing in situ measurements and corresponding divergence-free IMF derived from the Maxwell's equations. We use this model to describe the 27-day variation of cosmic rays over the course of three consecutive minimum epochs of solar activity with opposite solar magnetic polarities (1995-1997, A>0; 1985-1987, A<0 and 2007-2009, A<0).
4. We show that results of the proposed 3-D modeling of the 27-day variation of the GCR intensity for different polarity periods of the solar magnetic cycle are compatible with the neutron monitors data. In order to reach a compatibility of the theoretical modeling with the experimental data for the last minimum epoch of solar activity (2007-2009, A<0), a parallel diffusion coefficient was increased by ~ 40%.

ACKNOWLEDGMENTS
Authors thank investigators of the websites:
http://spidr.ngdc.noaa.gov/spidr/,
http://omniweb.gsfc.nasa.gov/index.html
http://cr0.izmiran.rssi.ru/kiel/main.htm
http://www.nmdb.eu

Authors thank Editor and the Referees for valuable remarks and suggestions helping us to improve the paper.

This work has been supported by Foundation for Polish Science. Authors would like to thank COSPAR Committee for partial sponsorship to attend 37[th] Cospar Scientific Assembly.